# Negative Thermal Expansion Coefficient of Graphene Measured by Raman Spectroscopy

*Duhee Yoon,[†] Young-Woo Son,[‡] Hyeonsik Cheong,[*,†]*

†Department of Physics, Sogang University, Seoul 121-742, Korea; ‡Korea Institute for Advanced Study, Seoul 130-722, Korea.

**AUTHOR EMAIL ADDRESS:** dhyoon@sogang.ac.kr; hand@kias.re.kr; hcheong@sogang.ac.kr

**CORRESPONDING AUTHOR FOOTNOTE:** E-mail: hcheong@sogang.ac.kr; TEL: 82-2-705-8434; Fax: 82-2-717-8434

**ABSTRACT**: The thermal expansion coefficient (TEC) of single-layer graphene is estimated with temperature-dependent Raman spectroscopy in the temperature range between 200 and 400 K. It is found to be strongly dependent on temperature but remains negative in the whole temperature range, with a room temperature value of $(-8.0 \pm 0.7) \times 10^{-6}$ K$^{-1}$. The strain caused by the TEC mismatch between graphene and the substrate plays a crucial role in determining the physical properties of graphene, and hence its effect must be accounted for in the interpretation of experimental data taken at cryogenic or elevated temperatures.

Graphene is attracting much interest due to potential applications as a next generation electronic material[1-3] as well as its unique physical properties.[4-6] In particular, its superior thermal and mechanical properties, including high thermal conductivity[7-11] and extremely high mechanical strength that exceeds 100 GPa,[12] make it a prime candidate material for heat control in high-density, high-speed integrated electronic devices. For such applications, knowledge of the thermal expansion coefficient (TEC) as a function of temperature is crucial. In order to determine the TEC of graphene directly, it would be necessary to measure a *free-standing* graphene sample. However, since most graphene samples are fabricated on substrates or over a trench held at the edges, such a direct measurement would be extremely challenging, if not impossible. Here, we demonstrate that the TEC can be estimated by monitoring the strain caused by the TEC mismatch between the graphene sample and the substrate whose TEC is known.

Graphite is known to have a negative TEC in the temperature range of 0–700 K.[13] For single-layer graphene (SLG), several authors have calculated the TEC using various methods.[14-16] Mounet *et al*. estimated the TEC of graphene as a function of temperature by using a first-principles calculation and predicted that graphene has a negative TEC at least up to 2500 K.[16] Bao *et al*. experimentally estimated the TEC in the temperature range of 300–400 K by monitoring the miniscule change in the sagging of a graphene piece suspended over a trench and estimated that it is negative only up to ~350 K.[17] It is not yet clear whether this discrepancy between theory and experimental data is caused by uncertainties in the accuracy of the experimental measurements, or limitations in the theoretical calculation. Since precise knowledge of the TEC in a wide temperature range is crucial in designing graphene-based devices and heat management systems, more precise measurements are needed.

Raman spectroscopy is a useful tool to investigate structural and electronic properties of graphene.[18] The softening of the Raman bands under tensile strain and splitting of the *G* and 2*D* bands under uniaxial tension have been reported.[19-24] Raman spectroscopic measurements were also used to estimate the thermal conductivity of suspended graphene by monitoring the Raman *G* band under illumination of a tightly focused laser beam.[7-10] When the temperature of a graphene sample fabricated on a SiO$_2$/Si substrate is raised, two effects should be considered: the temperature dependence of the phonon frequencies and the modification of the phonon dispersion due to the strain caused by the mismatch of the TEC's of the substrate and graphene. Since most graphene samples are fabricated on $SiO_2$ substrates or over a trench held at the edges, the pure effect of temperature change on the Raman spectrum cannot be measured directly and compared with the calculations which usually assume a free standing graphene.[25] The discrepancy between the experimentally measured Raman frequency shift and the theoretical prediction[25] can be reconciled by accounting for the TEC mismatch between the substrate and graphene.

In this Letter, we report an experimental estimation of the TEC of graphene in the temperature range of 200–400 K by analyzing the temperature-dependent shift of the Raman *G* band of SLG on $SiO_2$ and by careful exclusions of the substrate effects. The measured TEC's in the range are all negative unlike the previous measurement showing a negative-to-positive change of the TEC.[17] Moreover, the TEC exhibits a strong temperature dependence and its value at room temperature is $(-8.0 \pm 0.7) \times 10^{-6}$ $K^{-1}$. Below 200 K or above 400 K, the effects depending on the materials properties of the substrate such as buckling or slipping of graphene occurs, which obscures a clear determination of TEC of SLG. Our work calls for careful considerations on the TEC matching between graphene and the substrate in determining various intrinsic physical properties of graphene over a wide temperature range.

Graphene samples used in this work were prepared on silicon substrates covered with a 300-nm-thick $SiO_2$ layer by mechanical exfoliation of natural graphite flakes. The number of graphene layers was determined by inspecting the line shape of the Raman *2D* band.[26–28] Temperature-dependent Raman spectra of graphene and graphite were obtained while cooling and heating the samples in a microscope cryostat where the temperature could be controlled between 4.2 K and 475 K. The 514.5-nm line of an Ar ion laser was used as the excitation source, and a low power (< 0.3 mW) was used to avoid unintentional heating. A long-working-distance microscope objective (40×, 0.6 N.A.) was used to focus the laser beam onto the sample and collect the scattered light. The Raman scattered light signal was dispersed by a Jobin-Yvon Triax 550 spectrometer (1800 grooves/mm) and detected with a liquid-nitrogen-cooled CCD detector. The spectral resolution was ~0.7 cm$^{-1}$.

Figure 1(a) shows the frequency shifts of the Raman $G$ band ($\Delta\omega_G$) of SLG, bilayer graphene (BLG), and graphite samples as a function of temperature. The Raman peaks redshift as temperature rises and blueshift as temperature falls from room temperature. The Raman peak shift of SLG as a function of temperature is largest. We did not find any appreciable change in the line shape of the G band, which implies that the doping level does not vary appreciably.[29,30] The temperature-dependent Raman shift of free-standing graphene is commonly attributed to the thermal expansion of the lattice ($\Delta\omega_G^E$) and an anharmonic effect ($\Delta\omega_G^A$) which changes the phonon self-energy, as given by

$$\Delta\omega_G(T_m) = \Delta\omega_G^E(T_m) + \Delta\omega_G^A(T_m), \qquad (1)$$

where $T_m$ is the measured temperature of the sample. Previously, the phonon anhramonicity of graphene was theoretically studied with density-functional theory[25] and experimentally measured by the temperature-dependent Raman spectroscopy.[31,32] In Fig. 1(a), our experimental results on graphite agree well with the theory, but the data for SLG show significant discrepancies with the theoretical curve on free standing graphene.[25] Moreover, we found blueshifts of the $G$ band in SLG and BLG at temperatures over 400 K. SiO$_2$ has a positive TEC,[33] whereas graphene and graphite are known to have negative TEC's near room temperature.[13–17] As shown in Fig. 1(b), the SiO$_2$ layer expands (contracts) whereas the graphene sheet contracts (expands) as the temperature rises (falls). This TEC mismatch would induce a biaxial tensile or compressive strain on the graphene sample as temperature deviates from room temperature. When temperature rises further, graphene may slip on the surface of the substrate because the tensile strain increases significantly over the weak Van der Waals (VdW) force pinning graphene on the substrate. We interpret that slips occurred in our SLG and BLG samples at temperatures over 400K by noting the blueshifts of the measured $G$ bands at 400 K. In the case of graphite, since the weak VdW force is not expected to exert a coherent strain on a thick graphite sample, there is no significant strain effect in our experimental data, which explains why the data are in good agreement with the theory which does not take strain into account. In the case of a compressive strain (cooling), the graphene sheet may buckle, forming wrinkles or bubbles.[34,35] In our data for SLG, there seems to be a kink in the plot of $\Delta\omega_G(T_m)$ near 200 K. This is an indication of the initiation of buckling due to a compressive strain. From the data, we can assume that SLG is pinned on $SiO_2$ and experiences coherent strains due to the TEC mismatch as temperature varies between 200 and 400 K. It should be noted that of several samples that we measured, some samples showed smaller shifts of the $G$ band for the same temperature range, in which case the slip was not observed in the temperature range used. (See supporting information) This can be understood in the following way: when the VdW interaction between the graphene sample and the substrate is not strong enough, the strain between them would not be coherent. In such cases, the biaxial strain on the graphene sample would be smaller than the coherently strained case. Since the built-in strain is smaller for the same temperature change, the slip would occur at a higher temperature or would not occur at all. The data presented here represents the cases where the slip or buckling occurs at the smallest temperature changes. It is, therefore, safe to assume that the strain is coherent up to the point where the slip or buckling occurs. If there is some small slippage between the sample and the substrate in the temperature range of interest (200 – 400 K), the estimated TEC value would be smaller than the true value. Therefore, our estimate should be taken as a lower bound of the magnitude of the TEC.

When temperature varies, both usual thermal effects and strains induced by the TEC mismatch between the substrates and graphene must be considered simultaneously. Hence, the temperature-dependent frequency shifts of the Raman $G$ band $\Delta\omega_G(T_m)$ of graphene on a substrate should be written as

$$\Delta\omega_G(T_m) = \Delta\omega_G^E(T_m) + \Delta\omega_G^A(T_m) + \Delta\omega_G^S(T_m), \qquad (2)$$

where $\Delta\omega_G^S(T_m)$ is the effect of the strain $\varepsilon(T_m)$ due to the TEC mismatch. It can be expressed as

$$\begin{aligned}\Delta\omega_G^S(T_m) &= \beta \cdot \varepsilon(T_m) \\ &= \beta \cdot \int_{297K}^{T_m}\left(\alpha_{SiO_2}(T) - \alpha_{graphene}(T)\right)dT,\end{aligned} \qquad (3)$$

where $\beta$ is the biaxial strain coefficient of the $G$ band, $\alpha_{SiO_2}$ and $\alpha_{graphene}$ are the temperature-dependent TEC's of $SiO_2$ and graphene, respectively. To estimate the applied strains on graphene due to the TEC mismatch, we first consider the available TEC data of individual components of the system as a function of temperature [Fig. 2(a)]: the experimentally determined TEC's of $SiO_2$[33] and calculated ones of graphite and free standing graphene.[16] Assuming that graphene is pinned on the $SiO_2$ substrate throughout the whole temperature range, we can estimate the strain $\varepsilon(T_m)$ on graphene as a function of temperature [Fig. 2(b)]. The biaxial strain coefficient, $\beta = \partial\omega_G / \partial\varepsilon$, can be obtained from separate measurements of the uniaxial strain dependence of the $G$ band frequency. It has been estimated to be $-70 \pm 3 \text{ cm}^{-1} / \%$ at room temperature.[19,24] Figure 2(c) shows $\Delta\omega_G^S(T_m)$ for SLG calculated with Eq. (3). Finally, based on the estimations described above, one can determine the frequency shift of the $G$ band only due to the thermal effect by subtracting the strain effect from the experimental data, $\Delta\omega_G(T_m) - \Delta\omega_G^S(T_m)$. The result is shown in Fig. 2(d). There is still significant difference between the calculation and the experimental data near room temperature. Hence, we can conclude that the estimation of phonon frequency shifts based on the existing TEC estimation of graphene is not capable of reproducing the observed shifts of the Raman $G$ band in the wide range of temperature.

On the other hand, one can use the TEC of SLG as a fitting parameter instead and fit the experimental data to a recent theoretical calculation on the temperature dependence of the $G$-phonon frequency including phonon-phonon and electron-phonon interactions in *free-standing* graphene.[25] This approach can be justified because the calculation of the temperature dependence of the phonon frequency is usually more reliable than that of the TEC. Recent measurements on the temperature dependence of the phonon frequency of unsupported graphene[36] or nearly-freestanding graphene[37] were consistent with the calculations of Bonini *et al.*[25] As discussed earlier, within the temperature range between 200 and 400 K, graphene on $SiO_2$ can be regarded as coherently strained due to the TEC mismatch. Therefore, we take the temperature-dependent TEC of SLG in Eq. (3) as a fitting parameter of the frequency shifts to account for the strain effects. As shown in Fig. 3(a), when such strain effect on the shift $\Delta\omega_G^S(T_m)$ is subtracted from the measured shift $\Delta\omega_G(T_m)$, the experimental data are described well with the theoretical calculation of Bonini *et al*.[25] in the temperature range of 200–400 K. The TEC used to obtain the best fit is plotted in Fig. 3(b). The TEC at room temperature is estimated to be $\alpha_{SLG} \approx (-8.0 \pm 0.7) \times 10^{-6}$ K$^{-1}$, which shows a good agreement with the previous experimental value of $\alpha_{SLG} \approx -7 \times 10^{-6}$ K$^{-1}$.[16] However, our estimated TEC is always negative between 200 K and 400 K whereas the one obtained by Bao *et al*. exhibits a negative-to-positive transition at ~350 K.[17] It should be noted that Zakharchenko *et al*. calculated the negative thermal expansion coefficient of single layer graphene and found the negative-to positive transition to occur at ~ 900 K. They also estimated the *average* TEC between 0 and 300 K to be $\alpha_{SLG} \approx (-4.8 \pm 1.0) \times 10^{-6}$ K$^{-1}$, which is about half of our estimate at 300 K.

It should be noted here that in estimating $\Delta\omega_G^S(T_m)$, we assumed that the biaxial strain coefficient of the G band ($\beta$) is independent of temperature. Here $\beta = -2\omega_0\gamma$, where $\omega_0$ is the *G*-phonon frequency without strain and $\gamma$ is the Grüneisen parameter of SLG. At the moment, $\gamma$ has been measured only at room temperature.[19,24] Since the Grüneisen parameters for many other materials vary with temperature,[38] it is possible that the $\gamma$ value for SLG is also dependent on temperature. Since there is no experimental data or theoretical estimates of the temperature dependence of $\gamma$ of SLG, we assumed that its value is independent of temperature. This assumption introduces some uncertainty in our estimate of $\Delta\omega_G^S(T_m)$. For graphite, $\gamma$ decreases weakly with temperature above 200 K, and the variation is estimated to be less than 20% between 200 K and 300 K.[39] If we assume, as a first approximation, that $\gamma$ decreases linearly by 20% per 100 K from 200 K to 400 K, our result is modified as the double-dotted curve in Fig. 3(b). Though the overall temperature dependence of the TEC is reduced, the main features of the result, including the room temperature value of the TEC, are not affected significantly.

Our Raman experiments show that the large mismatch of TEC's between SiO$_2$ and SLG at low temperatures results in significant variations of the physical properties of SLG. As shown in Fig. 3(a), the Raman $G$ band deviates away from the theoretical estimations[25] below 200 K. Since the calculation assumes a coherent biaxial strain on SLG by the substrate, the data indicate occurrences of large mechanical distortions such as buckling at low temperatures. If buckling occurs, transport measurements may be severely modified due to the morphology. In such cases, the inhomogeneous strain may result in a pseudo-magnetic field,[34,40] which would affect many transport properties at low temperatures. Hence, the effects of possible strain due to the TEC mismatch between the SLG sample and the substrate, which have been by and large ignored so far, should be considered carefully.

In conclusion, we estimated the temperature dependence of the TEC of SLG with temperature-dependent Raman spectroscopy in the temperature range between 200 and 400 K. It is found to be negative in the whole range, in contradiction to a previous estimate[17]; and varies strongly with temperature, with a room temperature value of $(-8.0 \pm 0.7) \times 10^{-6}$ K$^{-1}$. We show that the effects of strain caused by the TEC mismatch between SLG and the substrate must be considered in interpreting the data from low-temperature transport measurements.

**ACKNOWLEDGMENT:** This work was supported by Mid-career Researcher Program through NRF grant funded by the MEST (No. 2011-0017605). Y.-W. S. was supported in part by the NRF grant funded by MEST (Quantum Metamaterials Research Center, R11-2008-053-01002-0 and Nano R&D program 2008-03670).

**FIGURE CAPTIONS**

**Figure 1.** (a) Raman frequency shifts of single-layer graphene (SLG), bilayer graphene (BLG), and graphite as a function of temperature. The solid and dashed lines are calculated results by Bonini *et al*. for SLG and graphite, respectively.[25] (b) Thermal expansion and contraction of graphene on a substrate (SiO$_2$) in cooling and heating processes.

**Figure 2.** (a) Measured TEC ($\alpha$) of SiO$_2$[33] and calculated $\alpha$ for graphite and graphene.[16] (b) Strain $\varepsilon$ on graphene due to TEC mismatch. (c) Estimated frequency shift of the $G$ band induced by strain, $\Delta\omega_G^S$. (d)

Temperature dependence of the $G$ band Raman frequency of SLG corrected for the strain effect, $\Delta\omega_G - \Delta\omega_G^S$. The solid curve is a theoretical estimate.[25]

**Figure 3.** (a) Temperature dependence of the $G$ band Raman frequency of freestanding SLG obtained by correcting for the strain effect, $\Delta\omega_G - \Delta\omega_G^S$, using the TEC as the fitting parameter. The solid curve is a theoretical estimate by Bonini *et al.*[25] (b) (Solid curve) TEC ($\alpha$) of SLG that gives the best fit between the data and the theoretical estimate in (a); (broken curves) theoretical calculation by Mounet *et al.*[16] and previous experimental estimate by Bao *et al.*[17]; (double-dotted curve) TEC obtained when $\gamma$ is assumed to vary 20% (see text).

**FIGURES**

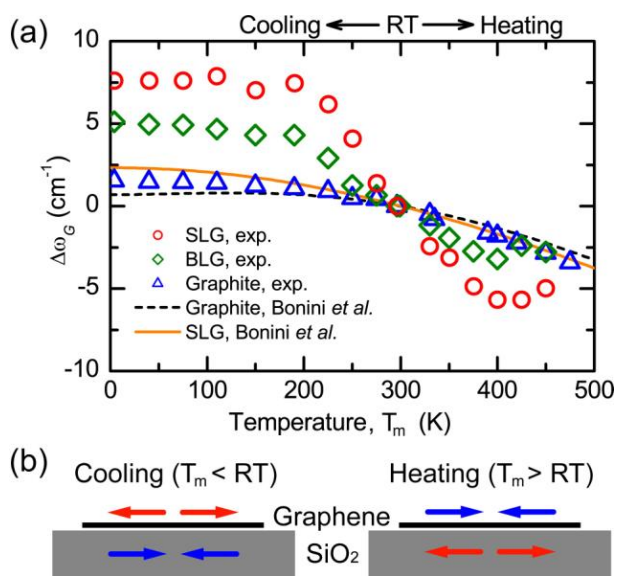

FIGURE 1. Yoon *et al.*

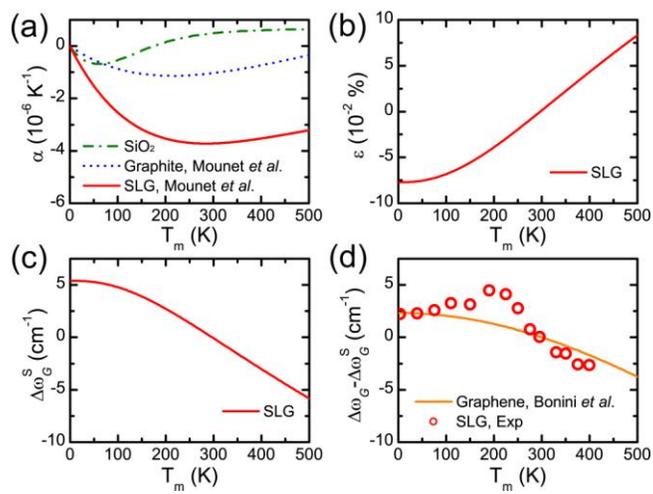

**FIGURE 2. Yoon *et al.***

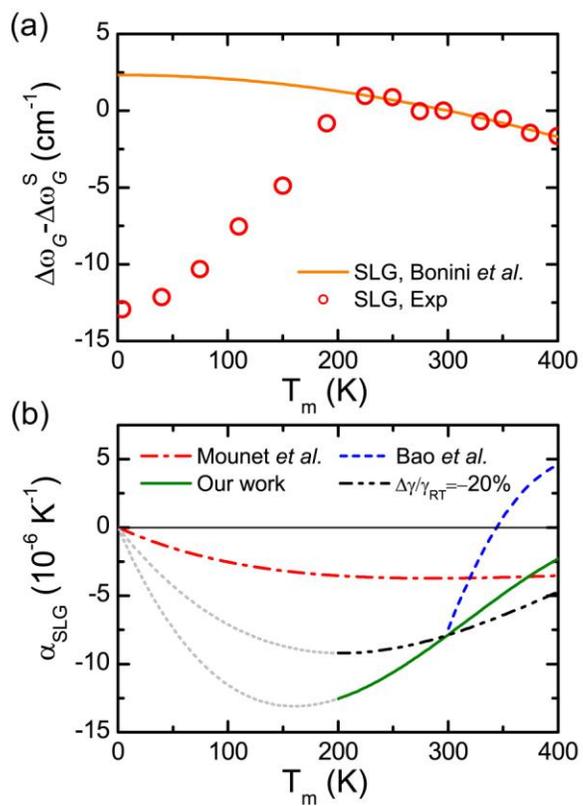

**FIGURE 3. Yoon *et al.***

**SYNOPSIS TOC**

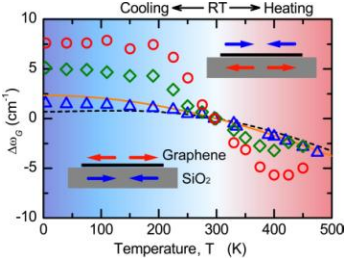

# Negative Thermal Expansion Coefficient of Graphene

# Measured by Raman Spectroscopy

*Duhee Yoon,[†] Young-Woo Son,[‡] Hyeonsik Cheong,[\*,†]*

[†]Department of Physics, Sogang University, Seoul 121-742, Korea; [‡]Korea Institute for Advanced Study,

Seoul 130-722, Korea.

**AUTHOR EMAIL ADDRESS:** dhyoon@sogang.ac.kr; hand@kias.re.kr; hcheong@sogang.ac.kr

**CORRESPONDING AUTHOR FOOTNOTE:** E-mail: hcheong@sogang.ac.kr; TEL: 82-2-705-8434;

Fax: 82-2-717-8434

## 1. Pictures of graphene samples

Several samples were used in this study. Figure 1 shows some representative samples. The number of layers was estimated by the shape of the 2*D* band in the Raman spectra (Fig. S1d).

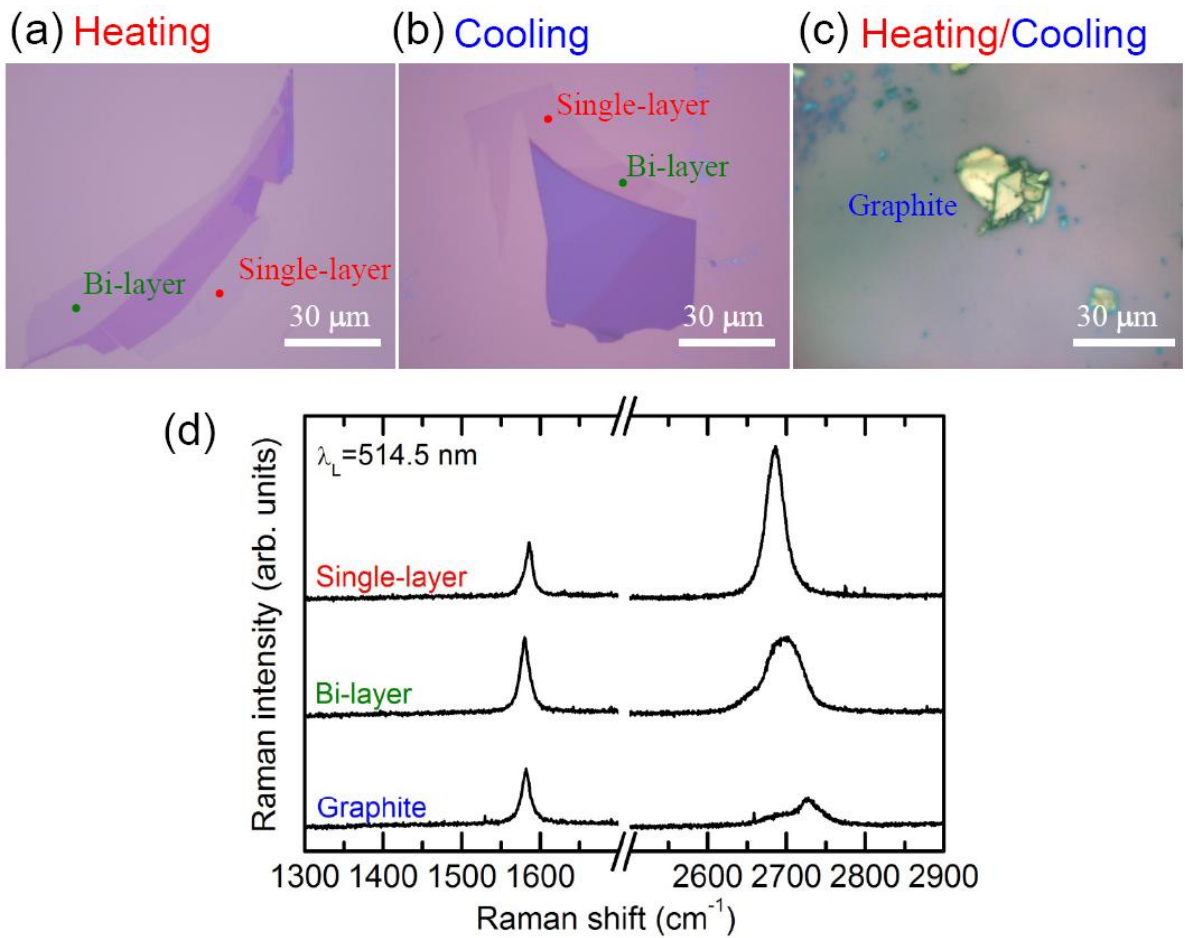

**Figure S1.** (a) and (b) Optical microscope images of graphene samples used in the study. The single- and bi-layer regions are indicated. (c) The graphite sample. (d) The Raman spectra of the single- and bi-layer regions of sample shown in (a) and the graphite sample in (c).

## 2. Hysteresis of Raman *G* band of single-layer graphene as a function of temperature

The hysteretic behavior of the Raman spectrum of a single-layer graphene sample was examined to confirm our interpretation of slipping. As shown in Fig. S2, the *G* band shows no hysteresis when the sample is heated only to 375 K and cooled back to 300 K. However, when it is heated further, a slip occurred between 375 K and 420 K. When it is cooled back to room temperature, the *G* band did not return to the original frequency. Furthermore, the *G* band exhibits a splitting and a blueshift, which suggests that the sample is under compressive strain. Under optical microscope, no discernible change was noticed (Fig. S2c).

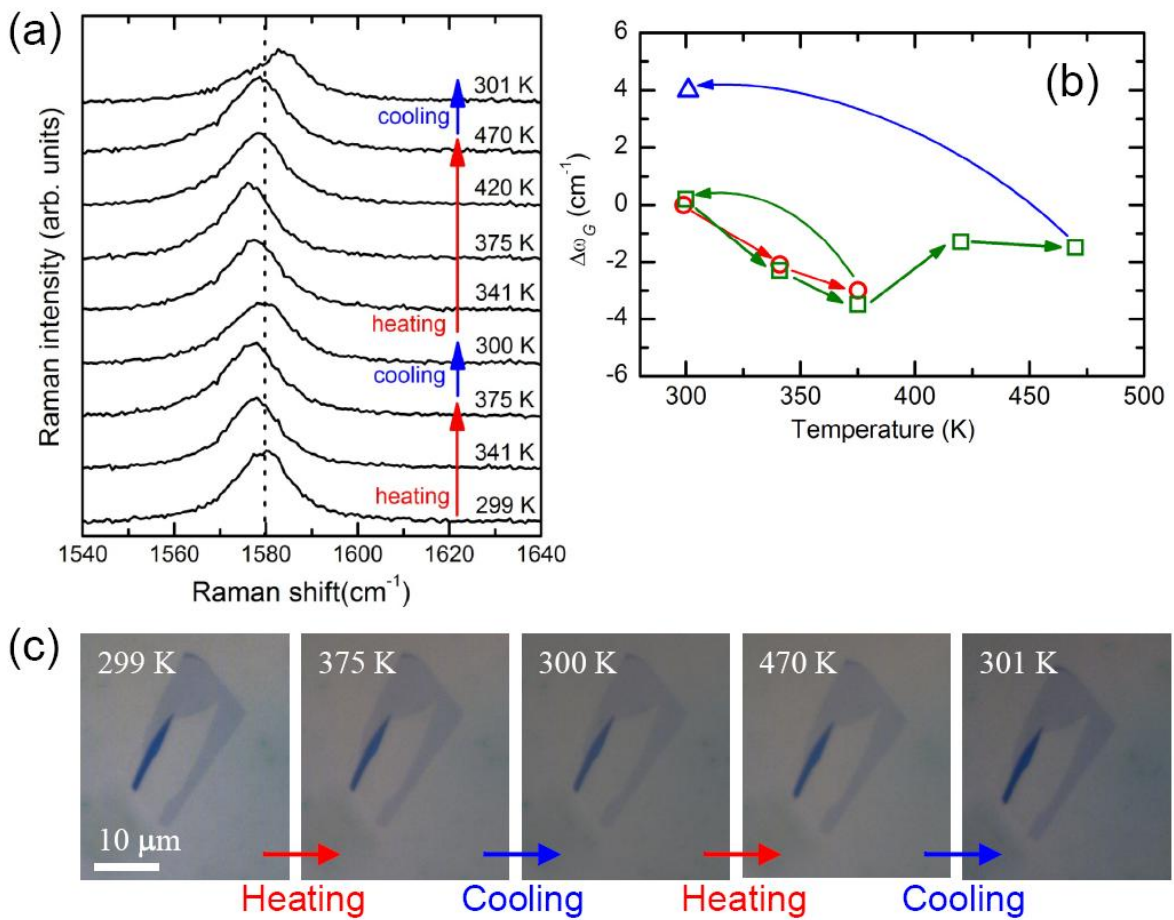

**Figure S2.** (a) Raman *G* band spectra of a single-layer graphene sample as the temperature is cycled. (b) Frequency shift of the Raman *G* band as a function of temperature. Between 300 K and 375 K, the temperature induced shift is reversible, whereas a large hysteresis is observed after the sample is heated beyond a 'slip'. (c) Optical microscope images of the sample at different stages of the temperature cycle.

## 3. Comparison between experimental results of various graphene samples

We measured several graphene samples. Some samples showed smaller shifts of the *G* band for the same temperature range, in which case the slip was not observed in the temperature range used. We used experimental data which has the largest frequency shift of the *G* band to estimate the TEC of graphene since this graphene sheet is well pined to the substrate until a slippage occurs.

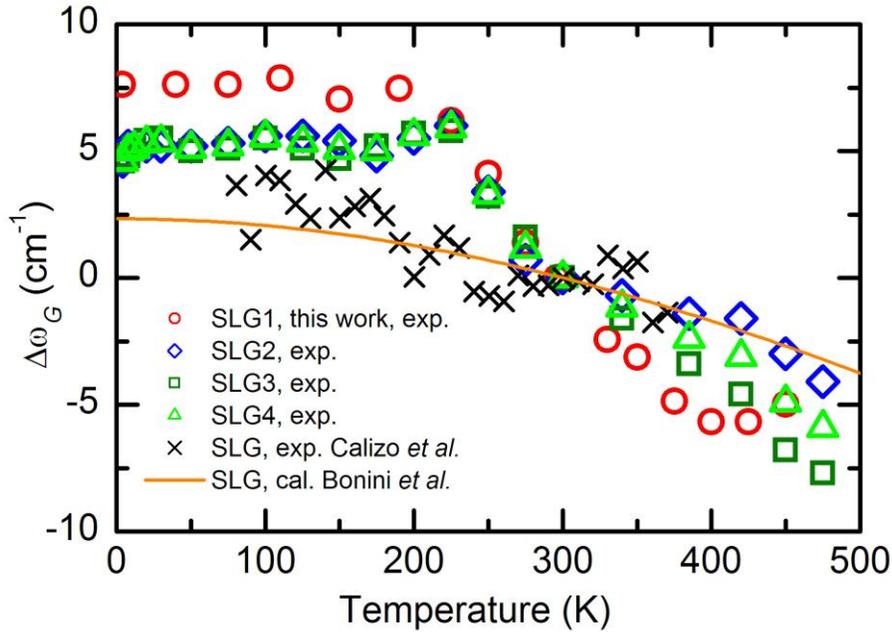

**Figure S3.** Raman frequency shifts of various single-layer graphene (SLG) samples. Cross marks are experimental results by Calizo *et al.* for SLG on SiO$_2$/Si substrate.[1] The solid lines are calculated results by Bonini *et al.* for *freestanding* SLG.[2]